\documentclass[10pt,draft,reqno]{amsart}
     \makeatletter
     \def\section{\@startsection{section}{1}%
     \z@{.7\linespacing\@plus\linespacing}{.5\linespacing}%
     {\bfseries
     \centering
     }}
     \def\@secnumfont{\bfseries}
     \makeatother
\newtheorem{theorem}{Theorem}[section]

\theoremstyle{definition}

\theoremstyle{remark}

\numberwithin{equation}{section} 
\newcommand{\abs}[1]{\lvert#1\rvert}
\begin{document}

\title[Continuous Quantum Games]{von Neumann's Minimax Theorem for Continuous Quantum Games}

\author{Luigi Accardi}
\address{ Centro Vito Volterra, Universit\`{a} di Roma Tor Vergata, via
Columbia  2, 00133 Roma, Italy}
\email{accardi@volterra.mat.uniroma2.it}

\author[Andreas Boukas]{Andreas Boukas*}
\thanks{* Corresponding author}
\address{Centro Vito Volterra, Universit\`{a} di Roma Tor Vergata, via
Columbia  2, 00133 Roma, Italy and Hellenic Open University,
Graduate School of Mathematics, Patras,  26335, Greece}
\email{boukas.andreas@ac.eap.gr}



\begin{abstract}
The concept of a classical player, corresponding to a classical
random variable, is extended to include quantum random variables
in the form of self adjoint operators on infinite dimensional
Hilbert space. A quantum version of Von Neumann's Minimax theorem
for infinite dimensional (or continuous) games is proved.
\end{abstract}

\maketitle

\section{Introduction: Classical two-person zero-sum games}\label{11}

\noindent In classical \textit{zero-sum infinite-dimensional  (or
continuous)  games} between two players, called Blue and Red, each
player has an infinite number of \textit{moves} (or \textit{pure
strategies}) available in each play of the game. The moves of Blue
and Red are identified \cite{Dre81} with the points of some closed
and bounded intervals $\lbrack a,b \rbrack \subset \mathbb{R}$ and
$\lbrack c, d \rbrack \subset \mathbb{R}$ respectively. To the
case when Blue makes choice $\lambda \in \lbrack a,b \rbrack $ and
Red makes choice $l \in \lbrack c,d \rbrack$ we assign a numerical
non-negative \textit{payoff} $Z(\lambda,l)$ to Blue and a
corresponding payoff $-Z(\lambda,l)$ to Red. We assume that both
players take a conservative approach, in the sense that Blue wants
to maximize his minimum payoff and Red wants to minimize the
maximum payoff to Blue.  If there exists $( \lambda_0,l_0 ) \in
\lbrack a,b \rbrack  \times \lbrack c,d \rbrack$ such that
\[
\max_\lambda \, \min_l \, Z(\lambda,l)=Z(\lambda_0,l_0)=\min_l
\,\max_\lambda \,Z(\lambda,l)
\]
\noindent then $(\lambda_0,l_0)$ is a \textit{saddle point} of $Z$ and
$\lambda_0,l_0$  are \textit{optimal moves}  for Blue and Red
respectively. If $Z$ has no saddle point then Blue and Red must
use \textit{mixed strategies} i.e, see \cite{Dre81}, they must
alter their moves and choose them using random devises represented
by classical probability distribution functions
\[
 F:\lbrack a,b \rbrack \to
\lbrack 0,1 \rbrack \,,\, G: \lbrack c,d \rbrack \to \lbrack 0,1
\rbrack
\]
\noindent associated with the probability measure spaces
$(\Omega_1,\sigma_1,\mu_1 )$ and $(\Omega_2,\sigma_2,\mu_2 )$,
respectively, that describe the outcomes of the random devices
used, respectively, by Blue and Red. Here, for $i=1, 2$, the
$\Omega_i$'s  are the sample spaces,   the $\mu_i $'s  are the
probability measures on the $\Omega_i$ 's , and the $\sigma_i$'s
are the corresponding $\sigma$-algebras of measurable subsets of
the $\Omega_i$'s. We may then think of the \textit{classical
players} Blue and Red as random variables $B$ and $R$, i.e., as
measurable functions
\begin{equation}\label{rv}
B:(\Omega_1,\sigma_1,\mu_1 )\to \lbrack a,b
\rbrack\,,\,R:(\Omega_2,\sigma_2,\mu_2)\to \lbrack c,d \rbrack \ .
\end{equation}
\noindent Then
\begin{equation}
F (\lambda )=\mu_1 \left(\{\omega_1\in \Omega_1: B(\omega_1 )\le
\lambda \}\right)= \Pr\, ( \mbox{ player } B \mbox{ makes a move }
\le \lambda )\label{prob1}
\end{equation}
\noindent and
\begin{equation}
 G(l)=\mu_2\left(\{\omega_2 \in
\Omega_2:R(\omega_2 )\le l \}\right)= \Pr \,(\mbox{ player } R
\mbox{ makes a move } \le l )\ .
\end{equation}
\noindent If we let the double Riemann-Stieltjes integral
\[
K(F,G )=\int_c^d\,\int_a^b\,Z(\lambda,l)\,dF(\lambda)\,dG(l)
\]
\noindent be the \textit{total expected payoff to $B$}, then the fundamental
problem of two-person zero-sum continuous game theory is the
existence of probability distributions, i.e., of mixed strategies,
$F^*$ and $G^*$ such that
\begin{equation}\label{val}
\max_F\,\min_G\,K(F,G )=K(F^*,G^* )=\min_G\,\max_F\,K(F,G) \  .
\end{equation}
\noindent If such $F^*$ and $G^*$ exist, then  $K(F^*,G^*)$ is the
\textit{value} of the game. If the payoff function $Z$ is
continuous then,   by the extension of von Neumann's Minimax
Theorem \cite{vn}  to infinite dimensional games \cite{Dre81,
Rand}, such $F^*$ and $G^*$ always exist. For a historical study
of von Neumann's Minimax Theorem  we refer to \cite{His}.

\medskip

\noindent Modern proofs of the existence of optimal strategies typically use
Kakutani's fixed point theorem \cite{Ka} for strategies in
$\mathbb{R}^n$ or, in the case of infinite dimensional sets of
strategies, its extensions to Banach or locally convex topological
vector spaces \cite{Glick, Fan}. In all cases, a compact and
convex set of strategies from which to choose, is required.

\medskip

\noindent The remaining sections are structured as follows:

\medskip

\noindent In Section \ref{22} we describe how the classical concept of a
\textit{player} can be translated into the language of quantum
mechanics \cite{Ho, Fein, Hall, vPa92}, i.e., in terms of
self-adjoint operators on Hilbert space, also referred to as
\textit{observables} or \textit{quantum random variables}, due to
the fact that their spectrum is real. In Section \ref{3} we
describe Kakutani's fixed point theorem for Banach spaces, in the
form that we are going to use it \cite{Ka, Fan, Glick, Smart}. In
Section \ref{4} we describe a class of compact and convex sets of
positive operators of trace one, that will serve as our sets of
mixed quantum strategies available to the two players. In Section
\ref{5} we describe how the classical two-person zero-sum game
setup can be formulated in terms of: self-adjoint operators on
Hilbert space, the spectral theorem, and positive operators of
unit trace. We then prove the quantum version of the Minimax
Theorem, i.e., Theorem \ref{QMT}. We remark that finite
dimensional two-person zero-sum quantum games were considered in
\cite{Bou}.

 \section{From Classical to Quantum Players}\label{22}

\noindent For a probability measure space $(\Omega,\sigma,\mu)$ as in
Section \ref{11},  $L^2(\Omega,\sigma,\mu)$, denotes the Hilbert
space of all $\mu$- equivalence classes of square-integrable
complex-valued functions $f$ defined on $\Omega$ with inner
product
\[
 \langle f, g \rangle =
 \int_{\Omega}\,\bar{f}(\omega)\,g(\omega)\,d\mu(\omega) \ .
 \]
 \noindent To pass from the classical to a quantum formulation of game theory we notice
that with the classical random variables, i.e., with the classical
players  $B$ and $R$ of (\ref{rv}), we can associate self-adjoint
multiplication operators
\[
\mathcal{B}:L^2(\Omega_1,\sigma_1,\mu_1)\rightarrow
L^2(\Omega_1,\sigma_1,\mu_1)\,,\,
\mathcal{R}:L^2(\Omega_2,\sigma_2,\mu_2)\rightarrow
L^2(\Omega_2,\sigma_2,\mu_2)
\]
\noindent defined pointwise by
\[
\mathcal{B}(f)(\omega_1)=B(\omega_1)\,f(\omega_1)\,,\,
\mathcal{R}(g)(\omega_2)=R(\omega_2)\,g(\omega_2) \ .
\]

 \medskip

\noindent In general, for $f\in L^2 (\Omega,\sigma,\mu )$ we let
$\rho=|f\rangle \langle f|$ denote the operator
\[
 \rho:L^2 (\Omega,\sigma,\mu )\rightarrow L^2
(\Omega,\sigma,\mu )
\]
\noindent defined by
\[
\rho (g)= |f \rangle \langle f| (g)=\langle f,g \rangle f
\]
\noindent  and for $\lambda \in \mathbb{R}$
we let $E(\lambda)$ denote the projection operator
\[
 E(\lambda) :L^2 (\Omega,\sigma,\mu )\rightarrow L^2
(\Omega,\sigma,\mu )
\]
\noindent defined by
\[
E(\lambda)(g)=\chi_{( -\infty, \lambda ]} g \ .
\]
\noindent If $f\equiv 1$ then $\rho=|f\rangle \langle f|$ is a
\textit{state} i.e a positive operator of unit trace and, in
analogy to (\ref{prob1}),
\[
F(\lambda):= {\rm tr}\,\rho E(\lambda) =\langle1,  E
(\lambda)1\rangle=\mu \left( ( -\infty,\lambda ]\right)=\Pr
\,(\mbox{player} B \mbox{ makes a move} \le\lambda)\  .
\]

\noindent  It is then suggested that we think of a \textit{quantum player} as a
self-adjoint operator
\[
T=\int_{ \mathbb{R}} \,\lambda \,\,dE (\lambda)
\]
\noindent on some infinite dimensional separable Hilbert space
$\mathcal{H}$, whose available moves coincide with its spectrum
$\sigma(T)$,  with the projection $E (\lambda)$ interpreted as the
event \textit{player $T$ makes a move $\le\lambda$}, and with
probability distribution
\[
F: \lambda \in \mathbb{R}\to F(\lambda)={\rm tr}\,\rho
E(\lambda)\in \lbrack 0, 1 \rbrack  \  ,
\]
\noindent determined by a state $\rho$ on $\mathcal{H}$.  We denote by $m(T)$
and $M(T)$ the \textit{lower bound} and \textit{upper bound} of
$T$, respectively, defined by
\[
m(T)=\inf_{\|x\|=1}\langle  T x,
x\rangle\,,\,M(T)=\sup_{\|x\|=1}\langle T x, x\rangle \ .
\]
\noindent If $T$ is bounded then $m(T)$ and $M(T)$ are finite and the
spectrum $\sigma(T)$ is contained in the interval $I_T=\lbrack
m(T), M(T)\rbrack$. In particular $m(T), M(T) \in \sigma(T)$. We
recall that $\sigma(T)$ is closed in $\mathbb{R}$.

\section{Kakutani's Fixed Point theorem}\label{3}

\noindent Following \cite{Smart}, if $\mathcal{S}$ is a subset of a normed
space $\mathcal{V}$ then a set-valued mapping $U:\mathcal{S}\to
P(\mathcal{S})$, where $P(\mathcal{S})$ is the \textit{power set}
of $\mathcal{S}$, is a $K$-\textit{mapping of $\mathcal{S}$ into
itself} if:

\medskip

\noindent (i) for each $x$ in $\mathcal{S}$, $U(x)$ is a compact
convex non-empty subset of $\mathcal{S}$;

\noindent (ii) the \textit{graph} of $U$, $\mathcal{G}(U)=\{(x,y):
y\in U(x)\}$ is closed in $\mathcal{S}\times \mathcal{S}$.

\medskip

\noindent Condition (ii) is equivalent to the following \textit{upper
semi-continuity} condition:

\medskip

\noindent (iii) if $x_n\to x$ in $\mathcal{S}$, $y_n\in U(x_n)$
and $y_n\to y$ then $y\in U(x)$.

\medskip

\noindent A \textit{fixed point} of  a $K$-\textit{mapping} $U$ is a point
$x\in \mathcal{S}$ such that $x\in U(x)$. A subset $\mathcal{S}$
 of a normed space $\mathcal{V}$ has \textit{the Kakutani property}
if each $K$-mapping $U$ of $\mathcal{S}$ into $\mathcal{S}$ has a
fixed point. Kakutani's  fixed point theorem \cite{Ka}  states
that every compact convex nonempty subset of $\mathbb{R}^n$ has
the Kakutani property. The theorem was extended in \cite{BK} from
$\mathbb{R}^n$ to any Banach space $\mathcal{V}$.

\section{A Compact Set of Quantum States}\label{4}

\noindent For an infinite dimensional separable Hilbert space $\mathcal{H}$
we denote by $\mathcal{B}( \mathcal{H})$ the Banach algebra of
bounded linear operators $A: \mathcal{H} \to \mathcal{H}$, with
the usual operator norm $\|A\|$, and by $\mathcal{T}(\mathcal{H})$
the Banach space of \textit{trace class} operators $T: \mathcal{H}
\to \mathcal{H}$ with the \textit{trace norm}
\[
\|T\|_1={\rm Tr}\,|T|={\rm Tr}\,(\sqrt{T^* T})=\sum_{i=1}^\infty
\langle e_i, \sqrt{T^* T} e_i \rangle \ ,
\]
\noindent where $\left( e_i \right)_{i=1}^\infty$ is any orthonormal basis
of $\mathcal{H}$. For $T\in\mathcal{T}(\mathcal{H})$ and
$A\in\mathcal{B}( \mathcal{H})$,
\[
|{\rm Tr}(TA)|\leq \|T\|_1 \|A\|\  .
\]
\noindent We denote by $\mathcal{S}(\mathcal{H})$ the closed convex subset
of $\mathcal{T}(\mathcal{H})$ consisting of all \textit{density
operators} (or \textit{quantum states}) in $\mathcal{H}$, i.e.,
all positive operators $\rho : \mathcal{H}\to \mathcal{H}$ with
${\rm Tr} (\rho)=1$. Equipped with the metric
\[
d\left(\rho_1, \rho_2\right)=\| \rho_1- \rho_2\|_1
\]
\noindent the \textit{state space}  $\mathcal{S}(\mathcal{H})$ is a complete
separable metric space and is a Banach space under $\| \cdot
\|_1$. In particular, $\mathcal{S}(\mathcal{H})\times
\mathcal{S}(\mathcal{H})$ is a Banach space under $\|(\rho_1 ,
 \rho_2)\|=\| \rho_1 \|_1+\| \rho_2 \|_1$. Unlike the finite
 dimensional case, $\mathcal{S}(\mathcal{H})$ is not compact.

\medskip

\noindent As shown in \cite{Ho}, if $\left( e_i \right)_{i=1}^{\infty}$ is
an orthonormal basis of $\mathcal{H}$, $\left( c_i
\right)_{i=1}^{\infty} \subseteq \mathbb{R}$ a sequence bounded
from below, and
\[
\mathcal{D}=\{\psi \in \mathcal{H}: \sum_{i=1}^\infty |c_i|^2
\langle e_i, \psi \rangle^2 <\infty \} \ ,
\]
\noindent then $\mathcal{D}$ is dense in $\mathcal{H}$  and the formula
\begin{equation}\label{en}
\mathcal{E} (\psi)=\sum_{i=1}^\infty c_i \langle e_i, \psi \rangle
e_i
\end{equation}
\noindent defines a self-adjoint operator $\mathcal{E}$
in $\mathcal{H}$ with domain $\mathcal{D}$. In particular,
$\mathcal{E}$ has the $e_i$'s as eigenvectors corresponding to its
eigenvalues $c_i$, $i=1, 2,...$ . If the multiplicities of the
$c_i$'s are finite and $c_i \to \infty$ as $i\to \infty$ then, by
Lemma 11.55 of \cite{Ho}, for an arbitrary positive constant $c$
the set
\begin{equation}\label{comp}
\mathcal{A}(c)=\{\rho \in \mathcal{S}(\mathcal{H}): {\rm Tr}\,
\rho \mathcal{E} \leq c\}
\end{equation}
\noindent is a compact subset of the metric space
$\mathcal{S}(\mathcal{H})$. By the linearity of the trace,
$\mathcal{A}(c)$ is also convex. Typically, $\mathcal{E}$ is the
\textit{energy operator} of a quantum oscillator system and we
think of $\mathcal{A}(c)$ as the set of \textit{quantum states
with mean energy $\leq c$}.

\section{The Quantum Minimax Theorem}\label{5}

\noindent Based on our concept of a quantum player described in Section
\ref{22}, we may set up a quantum game as follows:

\medskip

\noindent  In the notation of Section \ref{22}, for $i=1, 2$ let
$\mathcal{H}_i$ be  an infinite dimensional separable Hilbert
space, and let
\[
\mathcal{B}: \mathcal{H}_1\to \mathcal{H}_1 \,,\,\mathcal{R}:
\mathcal{H}_2\to \mathcal{H}_2
\]
\noindent be bounded self-adjoint operators with spectral resolutions
\[
\mathcal{B}=\int_{ \mathbb{R}} \,\lambda \,\,dE
(\lambda)\,,\,\mathcal{R}=\int_{ \mathbb{R}} \,l \,\,dE^{\prime}
(l)
\]
\noindent respectively. Let
\[
Z: I_{\mathcal{B}} \times I_{\mathcal{R}} \to \lbrack 0, +\infty )
\]
\noindent be continuous and not identically equal to zero on
$I_{\mathcal{B}} \times I_{\mathcal{R}}$. We could assume,
although we do not that here,  that $Z$ is equal to zero outside
$\sigma(\mathcal{B}) \times \sigma(\mathcal{R}) $, indicating the
impossibility of assigning a profit to non-observable moves.

\noindent For quantum states $\rho, \phi$  on $\mathcal{H}_1$ and
$\mathcal{H}_2$, respectively, for $\lambda, l \in \mathbb{R}$ we
define
\[
F_\rho(\lambda)={\rm tr}\,\rho E(\lambda)\,,\,G_{\phi}(l)={\rm
tr}\,\phi E^{\prime}(l) \  .
\]

\noindent With $\mathcal{B}$ and $\mathcal{R}$ we associate the total
expected payoff function
\[
K(\rho, \phi
)=\int_{I_{\mathcal{R}}}\,\int_{I_{\mathcal{B}}}\,Z(\lambda,l)\,dF_\rho(\lambda)\,dG_{\phi}(l)
\geq 0 \ .
\]

\noindent Finally, for given positive constants $c_1, c_2$, as in
(\ref{comp}),  we denote
\[
\mathcal{A}_1(c_1)=\{\rho \in \mathcal{S}(\mathcal{H}_1): {\rm
Tr}\, \rho \mathcal{E}_1 \leq c_1\}\,,\,\mathcal{A}_2(c_2)=\{\phi
\in \mathcal{S}(\mathcal{H}_2): {\rm Tr}\, \phi \mathcal{E}_2 \leq
c_2\}
\]
\noindent where  $\mathcal{E}_1, \mathcal{E}_1$  are of the type (\ref{en}).
We may now formulate and prove the following quantum version of
the Minimax Theorem which provides the quantum analogue of
(\ref{val}).

\begin{theorem}\label{QMT}  There exist quantum
states $\rho^*, {\phi}^*$ such that
\[
\max_{\rho \in \mathcal{A}_1(c_1)}\,\min_{\phi \in
\mathcal{A}_2(c_2)}\,K(\rho, \phi )=K(\rho^*, {\phi}^* )=\min_{
\phi \in \mathcal{A}_2(c_2)}\,\max_{\rho \in \mathcal{A}_1(c_1)
}\,K(\rho, \phi ) .
\]
\end{theorem}
\begin{proof} For each pair of quantum states $\rho$ and $\phi$, the set  $ I_{\mathcal{B}} \times
I_{\mathcal{R}}$ has finite measure
\[
 \left(F_\rho(M(\mathcal{B}))- F_\rho(m(\mathcal{B})  \right)\left(G_{\phi}( M(\mathcal{R} ))- G_{\phi}( m(\mathcal{R} )  \right)
\]
\noindent and $Z$ is continuous (thus measurable) and nonnegative on it.
Thus, by Fubini's theorem
\[
\int_{I_{\mathcal{R}}}\,\int_{I_{\mathcal{B}}}\,Z(\lambda,l)\,dF_\rho(\lambda)\,dG_{\phi}(l)=
\int_{I_{\mathcal{B}}}\,\int_{I_{\mathcal{R}}}\,\,Z(\lambda,l)\,dG_{\phi}(l)\,dF_\rho(\lambda)
\  .
\]
\noindent For each  $\rho \in \mathcal{A}_1(c_1)$, the mapping
\[
p_1: \phi \in \mathcal{A}_2(c_2) \to p_1(\phi
)=\int_{I_{\mathcal{B}}}\int_{I_{\mathcal{R}}}\,\,Z(\lambda,l)\,dG_{\phi}(l)\,dF_\rho(\lambda)
\in \mathbb{R}
\]
\noindent is continuous. To see that we notice that if $\left(\phi_n\right)$
is a sequence in $\mathcal{A}_2(c_2)$ with $\|\phi_n-\phi\|_1 \to
0$ as $n \to \infty$, where $\phi \in \mathcal{A}_2(c_2)$, then
\begin{align*}
\abs{p_1(\phi_n)-p_1(\phi)} =&\abs{
\int_{I_{\mathcal{B}}}\int_{I_{\mathcal{R}}}\,\,Z(\lambda,l)\,d\left(G_{\phi_n}-G_{\phi}\right)(l)\,dF_\rho(\lambda)
}\\
 \leq & V\left(G_{\phi_n}-G_{\phi}\right)V\left( F_\rho
\right) \, Z(\lambda_0,l_0)
\end{align*}
\noindent where
\[
 Z(\lambda_0,l_0)= \max_{\lambda \in
 I_{\mathcal{B}}, l \in
 I_{\mathcal{R}}}  Z(\lambda,l)
\]
\noindent and
\begin{align*}
 V\left(G_{\phi_n}-G_{\phi}\right) =&\inf \{K: \sum_{i=0}^k \abs{
 \left(G_{\phi_n}-G_{\phi}\right)(x_i)-\left(G_{\phi_n}-G_{\phi}\right)(x_{i-1})}
 \leq K  \}\\
V\left( F_\rho \right) =&\inf \{K: \sum_{i=0}^N \abs{
 F_\rho(y_i)-F_\rho(y_{i-1})}
 \leq K  \}   \  ,
\end{align*}
\noindent where the inequality must hold for all partitions
\[
\{x_0<x_1<...<x_{i-1}<x_i<...<x_k \}, k\in \mathbb{N},
\]
\noindent of $ I_{\mathcal{R}}$ and
\[
\{y_0<y_1<...<y_{i-1}<y_i<...<y_N \}, N \in \mathbb{N},
\]
\noindent of $I_{\mathcal{B}}$, are the \textit{total variations} (see
\cite{Fr} and \cite{Di}) of $G_{\phi_n}-G_{\phi}$ and $F_\rho$
over $ I_{\mathcal{R}}$ and $ I_{\mathcal{B}}$ respectively.

\noindent We have:
\begin{align*} 
&\sum_{i=1}^k |(G_{\phi_n} - G_{\phi})(x_i) - (G_{\phi_n} - G_{\phi})(x_{i-1})| = \sum_{i=1}^k |\text{tr}((\phi_n - \phi) P_i)| \\
&\le \sum_{i=1}^k \text{tr}(|\phi_n - \phi| P_i)  
= \text{tr}\left(|\phi_n - \phi| \sum_{i=1}^k P_i\right) 
\le \text{tr}(|\phi_n - \phi|) 
= \|\phi_n - \phi\|_1 
\end{align*}

\noindent where $P_i = E(x_i) - E(x_{i-1})$. The first inequality follows from $|\text{tr}(AB)| \le \text{tr}(|A|B)$ for self-adjoint $A$ and positive $B$, and the final inequality follows from $\sum P_i \le I$. Taking the supremum over all partitions, we obtain:
$$V(G_{\phi_n} - G_{\phi}) \le \|\phi_n - \phi\|_1$$

\noindent Similarly
\[
V\left( F_\rho \right)\leq  \|\rho\|_1=1 \  .
\]
\noindent Thus
\[
\abs{p_1(\phi_n)-p_1(\phi)}    \leq  \|\phi_n-\phi\|_1\,\,
Z(\lambda_0,l_0)
\]
\noindent so  $p_1$ is continuous for each $\rho \in \mathcal{A}_1(c_1)$ .
Thus, since $\mathcal{A}_2(c_2)$ is compact, for each $\rho \in
\mathcal{A}_1(c_1)$ there exists a (not necessarily unique)
quantum state $\phi^* (\rho)\in \mathcal{A}_2(c_2)$ at which $p_1$
attains its minimum, i.e., for each $\rho \in \mathcal{A}_1(c_1)$,
\begin{equation}\label{1}
K(\rho, \phi^*(\rho) )=\min_{\phi \in \mathcal{A}_2(c_2)}\,K(\rho,
\phi ) \   ,
\end{equation}
\noindent Similarly, for each $\phi \in \mathcal{A}_2(c_2)$ there exists a
(not necessarily unique) quantum state
$\rho^*(\phi)\in\mathcal{A}_1(c_1)$ at which the mapping
\[
p_2: \rho \in \mathcal{A}_1(c_1)  \to
p_2(\rho)=\int_{I_{\mathcal{B}}}\int_{I_{\mathcal{R}}}\,\,Z(\lambda,l)\,dG_{\phi}(l)\,dF_\rho(\lambda)
\in \mathbb{R} \  ,
\]
\noindent attains its maximum, i.e., for each $\phi \in \mathcal{A}_2(c_2)$,
\begin{equation}\label{2}
K(\rho^*(\phi), \phi )=\max_{\rho \in \mathcal{A}_1(c_1)}\,K(\rho,
\phi ) \  .
\end{equation}
\noindent  On the compact and convex subset $\mathcal{A}_1(c_1) \times  \mathcal{A}_2(c_2)$ of
$\mathcal{S}(\mathcal{H}) \times\mathcal{S}(\mathcal{H})$, we
define the point-to-set mapping $U$ by
\[
U(\rho, \phi )=\{( \rho^*(\phi),  \phi^*(\rho)):  \mbox{ such that
(\ref{1}) and (\ref{2}) are true} \} \  .
\]
We will show that $U$ is a $K$-mapping:

\medskip

\noindent As shown above, $U(\rho, \phi )$ is non-empty. Let $\left( (
\rho_n^*(\phi), \phi_n^*(\rho)) \right)$ be a sequence in $U(\rho,
\phi )$ converging to an element $(x,y)$ of $\mathcal{A}_1(c_1)
\times \mathcal{A}_2(c_2)$ with respect to the product metric
$d\times d$ on $\mathcal{S}(\mathcal{H})\times
\mathcal{S}(\mathcal{H})$. Then, as $n\to +\infty$,
\[
\|\rho_n^*(\phi)-x\|_1 \to 0\,,\,\|\phi_n^*(\rho)-y  \|_1 \to 0
 \ .
\]
\noindent We will show that $(x,y)\in U(\rho, \phi )$, so $U(\rho, \phi )$
is a closed (thus compact) subset of $\mathcal{A}_1(c_1) \times
\mathcal{A}_2(c_2)$ . Equivalently, we will show that $x$ and $y$
have properties (\ref{2}) and (\ref{1}) respectively.

\medskip

\noindent Let $\epsilon >0$ be given, and let $n_0 \in \mathbb{N}$ be such
that, for all $n\geq n_0$,
\[
\|\phi_n^*(\rho)-y  \|_1< \frac{\epsilon}{Z(\lambda_0,l_0)} \, ,\,
\|\rho_{n}^*(\phi)-x \|_1< \frac{\epsilon}{Z(\lambda_0,l_0)}
\]
\noindent Then, as in the proof of continuity of $p_1$,
\begin{align*}
 K(\rho, y)\leq & |K(\rho, y)-K(\rho,
\phi_{n_0}^*(\rho))|+K(\rho, \phi_{n_0}^*(\rho))\\
\leq& Z(\lambda_0,l_0) \|\phi_n^*(\rho)-y  \|_1+K(\rho,
\phi_{n_0}^*(\rho))\\
< & \epsilon+K(\rho, f)
\end{align*}
\noindent for all $f \in \mathcal{A}_2(c_2)$, since $\phi_{n_0}^*(\rho)$ has
property (\ref{1}). By the arbitrariness of $\epsilon$, it follows
that
\[
K(\rho, y) \leq K(\rho, f)
\]
\noindent for all $f \in \mathcal{A}_2(c_2)$, so $y$ has property (\ref{1}).
Similarly,
\begin{align*}
 K(x, \phi)= &|K(x, \phi)-K(\rho_{n_0}^*(\phi),
\phi)+K(\rho_{n_0}^*(\phi), \phi)|\\
\geq &K(\rho_{n_0}^*(\phi), \phi)- |K(x,
\phi)-K(\rho_{n_0}^*(\phi),\phi)|\\
\geq& K(\rho_{n_0}^*(\phi), \phi)- Z(\lambda_0,l_0)
\|\rho_{n_0}^*(\phi)-x \|_1\\
> & K(r, \phi)-\epsilon
\end{align*}
\noindent for all $r \in \mathcal{A}_1(c_1)$, since $\rho_{n_0}^*(\phi)$ has
property (\ref{2}). By the arbitrariness of $\epsilon$, it follows
that
\[
K(x, \phi) \geq K(r, \phi)
\]
\noindent for all $r \in \mathcal{A}_1(c_1)$, so $x$ has property (\ref{2}).
Thus $U(\rho, \phi )$ is compact.

\medskip

\noindent To show that $U(\rho, \phi )$ is also convex, let $(
\rho_i^*(\phi), \phi_i^*(\rho)) \in U(\rho, \phi )$, $i=1, 2$, and
let $t \in \lbrack 0, 1 \rbrack $. Then
\[
t\,( \rho_1^*(\phi),  \phi_1^*(\rho))+(1-t) ( \rho_2^*(\phi),
\phi_2^*(\rho)) =( t\, \rho_{1}^*(\phi) +(1-t) \rho_{2}^*(\phi),
t\, \phi_{1}^*(\rho) +(1-t) \phi_{2}^*(\rho)  )  \  .
\]
\noindent By the linearity of the trace,
\begin{align*}
K(t\,\rho_{1}^*(\phi) +(1-t) \rho_{2}^*(\phi), \phi)=&
t\, K( \rho_{1}^*(\phi), \phi) +(1-t) K(\rho_{2}^*(\phi), \phi)\\
\geq& t\, K( \rho, \phi) +(1-t) K(\rho, \phi)\\
=& K( \rho , \phi)
\end{align*}
\noindent for all $\rho \in \mathcal{A}_1(c_1)$, so $t\,
\rho_{1}^*(\phi)+(1-t) \rho_{2}^*(\phi)$ has property (\ref{2}).
Similarly
\[
K(\rho, t\, \phi_{1}^*(\rho) +(1-t) \phi_{2}^*(\rho) )\leq K( \rho
, \phi)
\]
\noindent for all $\phi \in \mathcal{A}_2(c_2)$, so $t\, \phi_{1}^*(\rho)
+(1-t) \phi_{2}^*(\rho)$ has property (\ref{1}). Thus $U(\rho,
\phi )$ is  convex.

\medskip

\noindent To  show that $U$ satisfies the upper semi-continuity condition of
Section \ref{3}, let $\left((\rho_n, \phi_n) \right)$ be a
sequence in $\mathcal{A}_1(c_1) \times \mathcal{A}_2(c_2)$
converging to an element $(\rho, \phi)$ of $\mathcal{A}_1(c_1)
\times \mathcal{A}_2(c_2)$ with respect to the product metric
$d\times d$ on $\mathcal{S}(\mathcal{H})\times
\mathcal{S}(\mathcal{H})$, i.e. $\|  \rho_n - \rho\|_1\to 0$ and
$\|  \phi_n - \phi\|_1\to 0$ as $n\to +\infty$, and let
$\left(\rho^*_n(\phi_n), \phi^*_n(\rho_n) \right) \in U(\rho_n,
\phi_n)$, i.e. for all $(r, f)\in \mathcal{A}_1(c_1) \times
\mathcal{A}_2(c_2)$ we have
\[ K(\rho_n, \phi^*_n(\rho_n))\leq
K(\rho_n, f)\,,\,K(\rho^*_n(\phi_n), \phi_n)\geq K(r, \phi_n)\ ,
\]
\noindent with $\left(\rho^*_n(\phi_n), \phi^*_n(\rho_n) \right)$ converging
to an element $(\rho^*, \phi^*)$ of $\mathcal{A}_1(c_1) \times
\mathcal{A}_2(c_2)$ with respect to the product metric $d\times d$
on $\mathcal{S}(\mathcal{H})\times \mathcal{S}(\mathcal{H})$, i.e.
$\|  \rho^*_n(\phi_n) - \rho^*\|_1\to 0$ and $\|  \phi^*_n(\rho_n)
- \phi^*\|_1\to 0$ as $n\to +\infty$. We will show that $(\rho^*,
\phi^* )\in U(\rho, \phi )$, i.e. that $\rho^*=\rho^*(\phi)$ and
$\phi^*=\phi^*(\rho)$, meaning that
\[
K(\rho, \phi^*)\leq K(\rho, f)\,,\,K(\rho^*, \phi)\geq K(r, \phi)\
,
\]
\noindent for all $(r, f)\in \mathcal{A}_1(c_1) \times \mathcal{A}_2(c_2)$.

\medskip

\noindent So let $\epsilon >0$ be given, and let $n_0 \in \mathbb{N}$ be
such that, for all $n\geq n_0$,
\[
\|\phi_n^*(\rho_n)-\phi^*  \|_1<
\frac{\epsilon}{2\,Z(\lambda_0,l_0)} \, ,\, \|\rho_{n}-\rho \|_1<
\frac{\epsilon}{2\,Z(\lambda_0,l_0)}
\]
\noindent and
\[
\|\rho_n^*(\phi_n)-\rho^*  \|_1<
\frac{\epsilon}{2\,Z(\lambda_0,l_0)} \, ,\, \|\phi_{n}-\phi \|_1<
\frac{\epsilon}{2\,Z(\lambda_0,l_0)} \ .
\]
\noindent Then
\begin{align*}
K(\rho, \phi^*)\leq & |K(\rho, \phi^*)-K(\rho_{n_0},
\phi_{n_0}^*(\rho_{n_0})  )|+K(\rho_{n_0}, \phi_{n_0}^*(\rho_{n_0}))\\
< & \frac{\epsilon}{2}+K(\rho_{n_0}, f)\\
\leq & \frac{\epsilon}{2}+|K(\rho_{n_0}, f)-K(\rho, f)|+K(\rho,
f)\\
< & \frac{\epsilon}{2}+\frac{\epsilon}{2}+K(\rho, f)\\
=&\epsilon+K(\rho, f)\  ,
\end{align*}
\noindent so $ K(\rho, \phi^*)\leq K(\rho, f)$ for all $ f\in
\mathcal{A}_2(c_2)$. Similarly,
\begin{align*}
 K(\rho^*, \phi)\geq & K(\rho^*_{n_0}(\phi_{n_0}),
\phi_{n_0})-|K(\rho^*, \phi)-K(\rho^*_{n_0}(\phi_{n_0}),
\phi_{n_0})|\\
>&K(r, \phi_{n_0})- \frac{\epsilon}{2}\\
\geq &K(r, \phi)- |K(r, \phi_{n_0})-K(r, \phi)|-\frac{\epsilon}{2}\\
> &K(r, \phi)- \frac{\epsilon}{2}-\frac{\epsilon}{2}\\
=&K(r, \phi)-\epsilon \  ,
\end{align*}
\noindent so $K(\rho^*, \phi)\geq K(r, \phi)$ for all $ r\in
\mathcal{A}_1(c_1)$. Thus $U$ is upper semi-continuous and by
Kakutani's theorem $U$ has a fixed point $(\rho^*, \phi^*)\in
U(\rho^*, \phi^*)$, i.e. such that
\[
K(\rho, \phi^* )\leq K(\rho, \phi ) \leq K(\rho^*, \phi ) \  ,
\]
\noindent for all $(\rho, \phi )\in \mathcal{A}_1(c_1) \times
\mathcal{A}_2(c_2)$, which is equivalent to
\[
\max_{\rho \in \mathcal{A}_1(c_1)}\,\min_{\phi \in
\mathcal{A}_2(c_2)}\,K(\rho, \phi )=K(\rho^*, {\phi}^* )=\min_{
\phi \in \mathcal{A}_2(c_2)}\,\max_{\rho \in \mathcal{A}_1(c_1)
}\,K(\rho, \phi ) .
\]
\end{proof}

\bibliographystyle{amsplain}

\end{document}